\documentclass[11pt,twoside]{article}
\usepackage{asp2010}

\resetcounters

\begin{document}

\title{General Relativistic Explosion Models of Core-Collapse Supernovae}
\author{Bernhard M\"uller$^1$, Andreas Marek$^1$, Hans-Thomas Janka$^1$, and Harald Dimmelmeier
\affil{$^1$Max-Planck-Insitut f\"ur Astrophysik, Karl-Schwarzschild-Str. 1, 85748 Garching, Germany}}

\begin{abstract}
We present results from the first generation of multi-dimensional
general relativistic neutrino hydrodynamics simulations of
core-collapse supernovae. A comparison with models computed using
either the purely Newtonian approximation or the ``effective
gravitational potential'' approach reveals appreciable quantitative
differences in the heating conditions and the gravitational wave
spectra. Our results underscore the important role of general
relativity in the supernova problem (which appears to be on par with
other important factors such as the dimensionality and the equation of
state) both for our understanding of the explosion dynamics as well as
for predictions of observable signatures.
\end{abstract}

\section{Introduction}
A core-collapse supernova occurs at the end of the life of a massive
star when it has exhausted its nuclear fuel at its centre. The nuclear
ashes form an iron core which undergoes a gravitational collapse once
it reaches a mass of roughly $1.4 M_\odot$. When the core reaches
supranuclear densities, the collapse is halted due the stiffening of
the equation of state, and a shock wave is launched as the core
rebounds. The initial kinetic energy of the shock is quickly spent for
disintegrating the nuclei in the infalling material material into free
nucleons, and the shock is further weakened by the rapid emission of
several $10^{51} \ \mathrm{erg}$ in neutrinos from the post-shock
region during the neutronization burst. The stalled shock typically
hovers at a radius of $\sim 100 \ldots 200 \ \mathrm{km}$ -- possibly
for hundreds of milliseconds -- until it is revived, and expels the
envelope of the progenitor.

The nature of the explosion mechanism has been a subject of intense
research in computational astrophysics for several decades. Several
mechanisms have been proposed to explain the revival of the shock, the
most prominent one being the ``delayed neutrino-driven mechanism''
\citep{bethe_85,wilson_85}, which relies on the deposition of energy by neutrinos in the
``gain layer'' behind the shock to re-energize the shock. Except for a
special class of progenitors with an O-Ne-Mg core \citep{kitaura_06}, this
mechanism only works in concert with multi-dimensional hydrodynamic
effects such as convection and the so-called ``standing accretion
shock instability'' (SASI, \citealp{blondin_03,foglizzo_06,ohnishi_06}), which
both increase the efficiency of neutrino heating in the gain
region. While some of the most ambitious supernova simulations in
axisymmetry \citep{buras_06_b, marek_09,bruenn_10,suwa_10} indicate
that the neutrino-driven mechanism may indeed work, there are also
alternative scenarios such a magnetohydrodynamically-driven explosions
\citep{bk_70,akiyama_03,dessart_07_a}, acoustically-powered supernovae
(\citealp{burrows_06}; see however \citealp{quataert_08} for
criticism), and explosions triggered by a QCD phase transition
\citep{sagert_09}.

From a computational point of view, core-collapse supernovae present a
number of challenges: Not only are they inherently multi-dimensional
phenomena due to the operation of hydrodynamical instabilities, but in
order to accurately capture the crucial effects of neutrino cooling
and heating in the optically thick and thin regimes, the problem needs
to be treated within the framework of kinetic theory. As the direct
solution of the full Boltzmann equation for neutrinos is currently
only feasible under the assumption of spherical symmetry \citep{yamada_99,liebendoerfer_04}, a
variety of different approximation strategies for the neutrino
transport are used in the most sophisticated multi-dimensional
supernova simulations, including ``ray-by-ray'' transport combined
with variable Eddington factor techniques \citep{buras_06_a} or
flux-limited multi-group diffusion schemes \citep{bruenn_10},
multi-angle Boltzmann transport without
energy-bin coupling \citep{ott_08_a}, two-moment schemes with an analytic
closure \citep{obergaulinger_11}, and the isotropic diffusion source approximation
\citep{liebendoerfer_09}.

Moreover, \emph{general relativity} (GR) plays a major role in the
supernova problem due to the compactness of the proto-neutron star
$(GM/Rc^2 \sim 0.1 \ldots 0.2)$ and the occurrence of high velocities
($\sim 0.3 c$). A general relativistic treatment is also required for
precise predictions of the gravitational wave signal from
core-collapse supernovae.  However, relativistic supernova simulations
with up-to-date neutrino transport have long been limited to the case
of spherical symmetry \citep{bruenn_01,yamada_99,liebendoerfer_04}. In
order to study the effects of GR in the context of
multi-dimensional models, and in particular its impact on the
gravitational wave emission from the supernovae core, we have recently
\citep{mueller_10} introduced a generalization of the ray-by-ray-plus
variable Eddington factor method \citep{rampp_02,buras_06_a}. In this
paper, we present results from the first generation of axisymmetric
relativistic supernova simulations using our \textsc{Vertex-CoCoNuT}
code and summarize the results of a comparison with the Newtonian case
as well as with the pseudo-Newtonian ``effective potential''
approximation.

\section{Relativistic Variable Eddington Factor Method}
In our approach to neutrino transport in core-collapse supernovae, we
solve the equations of GR hydrodynamics in the formulation of
\citet{banyuls_97} and use the xCFC approximation of
\citet{cordero_09} for the space-time metric, which is particularly
suitable for our purpose because of its excellent stability properties
and high accuracy in the core-collapse case \citep{cordero_09}. The
hydrodynamics solver is based on the \textsc{CoCoNuT} code
\citep{dimmelmeier_02_a,dimmelmeier_04}, an implementation of a HRSC
scheme with PPM reconstruction \citep{colella_84} and second-order
Runge-Kutta time-stepping. Different from the original code
\citep{mueller_10}, we rely on an improved scheme for maintaining
total energy conservation and employ the relativistic HLLC Riemann
solver of \citet{mignone_05_a} to resolve contact discontinuities in
the convective post-shock region more accurately.

In order to capture the effects of neutrino heating and cooling the
equations of hydrodynamics need to be coupled with a kinetic equation
for the neutrino distribution function $f$. Following the approach
already used for the Newtonian version the \textsc{Vertex} code
\citep{rampp_02,buras_06_a}, we simplify the full six-dimensional
phase-space problem by considering the first two angular moments $J,
H, K, L, \ldots$ of the energy-dependent radiation intensity and by
requiring $f$ to be axially symmetric in momentum space (but not in
real space) around the unit radius vector (the ray-by-ray
approximation). The transport problem can thus be reduced to
conservation equations with source terms for neutrino number, energy
and momentum; e.g., the equation for the zeroth moment $J$ reads
\begin{eqnarray}
\label{eq:momeq_cfc_number_j}
\lefteqn{\frac{\partial \sqrt{\gamma} W \left({J}+v_r {H}\right)}{\partial t}
+\frac{\partial}{\partial r}
\left[
\left(W \frac{\alpha}{\phi^2} -\beta_r v_r\right) \sqrt{\gamma} {{H}}+
\left(W v_r \frac{\alpha}{\phi^2} -\beta_r\right) \sqrt{\gamma}{{J}}
\right]
-}
\\
&&
\varepsilon 
\frac{\partial}{\partial \varepsilon}
\left\{
W \sqrt{\gamma} {{J}}
\left[
\frac{1}{r}\left(\beta_r-\frac{\alpha v_r}{\phi^2}\right)
+2 \left(\beta_r-\frac{\alpha v_r}{\phi^2}\right) \frac{\partial \ln \phi}{\partial r}
-2 \frac{\partial \ln \phi}{\partial t}
\right]
+
\right.
\nonumber
\\
&&
W \sqrt{\gamma} {{H}}
\left[
v_r \left(\frac{\partial \beta_r \phi^2}{\partial r}-2 \frac{\partial \ln \phi}{\partial t}\right)
-\frac{\alpha}{\phi^2}\frac{\partial \ln \alpha W}{\partial r}
+\alpha W^2 \left(\beta_r\frac{\partial v_r}{\partial r}-\frac{\partial v_r}{\partial t}\right)
\right]
-
\nonumber
\\
&&
\left.
\sqrt{\gamma} {{K}}
\left[
\frac{\beta_r W}{r}-\frac{\partial \beta_r W}{\partial r}
+W v_r r \frac{\partial}{\partial r}\left(\frac{\alpha}{r \phi^2}\right)
+W^3 \left(\frac{\alpha}{\phi^2}\frac{\partial v_r}{\partial r} + v_r\frac{\partial v_r}{\partial t}\right)
\right]
\right\}
=
\alpha \sqrt{\gamma} {{C}}^{(0)}
\nonumber
\end{eqnarray}
in the adopted gauge, with $\alpha$ denoting the lapse function, $\phi$ the
conformal factor, $\beta_r$ the radial shift vector, $\gamma$
the determinant of the three-metric, $v_r$ the radial velocity, $W$
the Lorentz factor, and $C^{(0)}$ the zeroth moment of the collision
integral. A similar equation is solved for $H$, and the higher moments $K$
and $L$ that are required to close the system are provided by means of
a formal solution of a simplified ``model'' Boltzmann equation.  It
should be noted that we fully retain the energy-dependence of the
moment equations, and fully include Doppler shift, gravitational
redshift an energy redistribution by inelastic scattering
of neutrinos off nucleons, nuclei, electrons, and other neutrinos.

\section{Results}
\begin{figure}
\includegraphics[width=0.45 \linewidth]{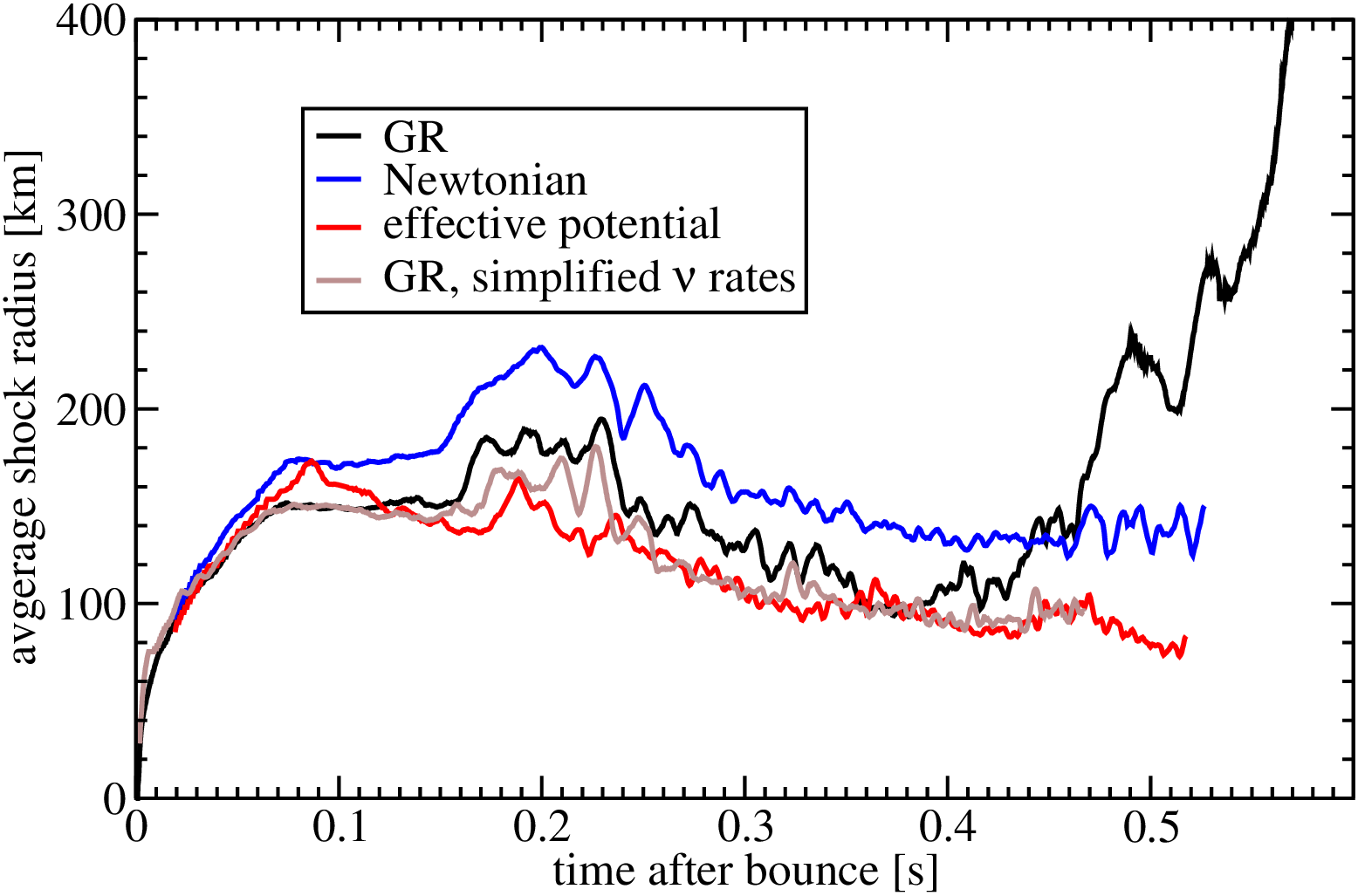}
\includegraphics[width=0.45 \linewidth]{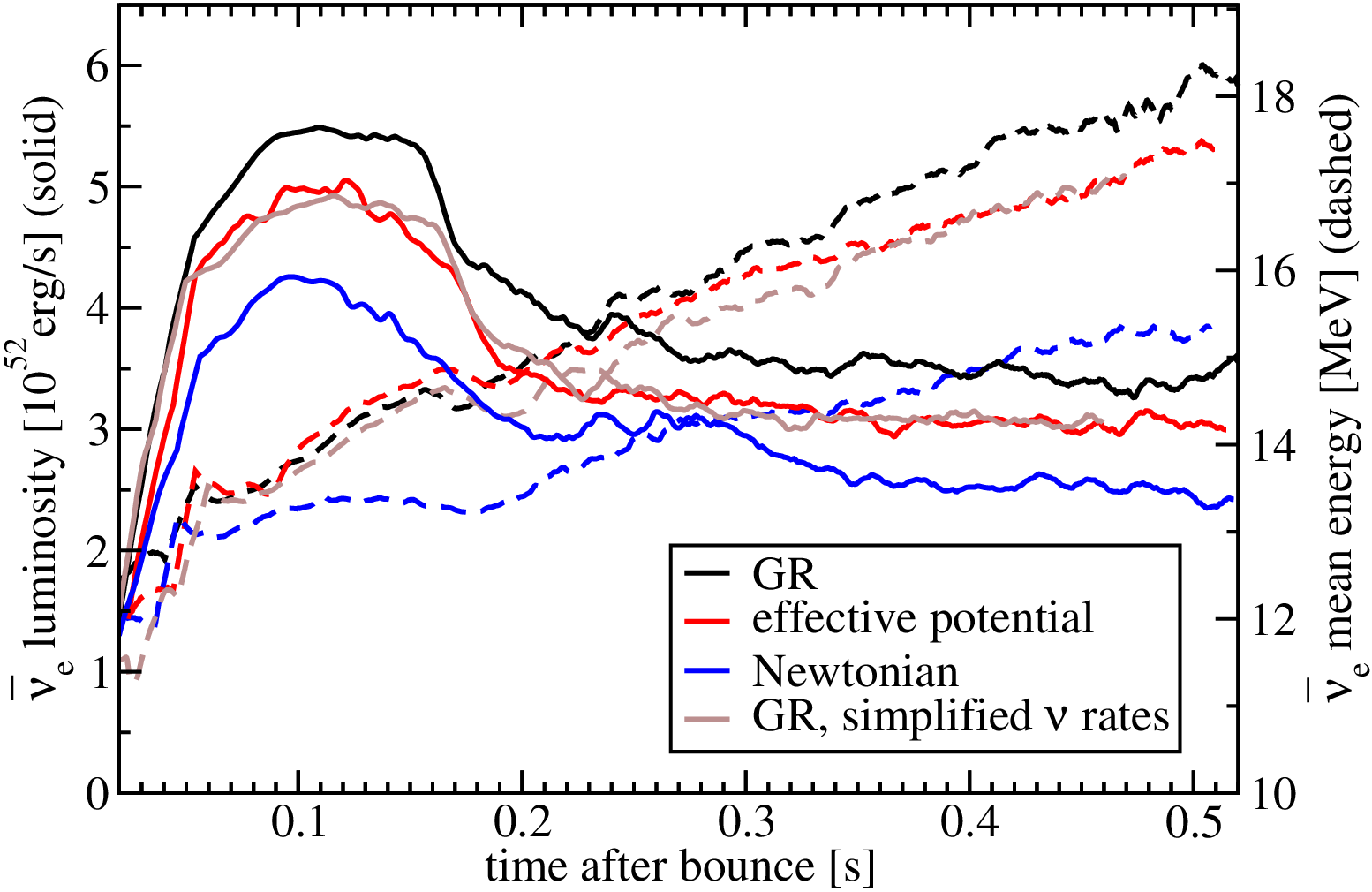}
\caption{Left: Evolution of the average shock radius for simulations
  of a $15 M_\odot$ progenitor with general relativistic hydrodynamics
  and neutrino transport (black), with an effective potential (red),
  in the purely Newtonian approximation (blue), and with a simplified
  set of neutrino interaction rates in the GR case (light brown).  Right:
  Electron antineutrino luminosities (solid) and mean energies
  (dashed) for the four different cases.
\label{fig:dynamics}
}
\end{figure}

Relativistic supernova simulations have been conducted for two
different non-rotating progenitors with $11.2 M_\odot$
\citep{woosley_02} and $15 M_\odot$ \citep{woosley_95}.  For the $15
M_\odot$ progenitor, additional models were computed to allow for a
comparison with the purely Newtonian case and the ``effective
potential'' approach \citep{marek_06} which mimics certain
strong-field effects by modifying the Newtonian gravitational
potential. Moreover, we also considered the case of slightly
simplified neutrino interaction rates, neglecting the effect of
recoil, high-density correlations and weak magnetism in
neutrino-nucleon reactions and ignoring reactions between different
neutrino flavours.  As the $15 M_\odot$ progenitor has already proved
to be a marginal case in earlier studies relying on the effective
potential approach \citep{marek_09}, it is ideally suited to
illustrate the impact of slightly different heating conditions
depending on the treatment of GR. However, the relativistic

Among the four $15 M_\odot$ simulations, only the GR model with the
best currently available set of neutrino opacities for our code
develops an explosion around $400 \ \mathrm{ms}$ after bounce (left
panel of Fig.~\ref{fig:dynamics}). The more optimistic evolution of
the GR model compared to the Newtonian run has been traced to slightly
higher surface temperatures of the more compact proto-neutron star,
which result in higher neutrino luminosities and mean energies (right
panel of Fig.~\ref{fig:dynamics}) and hence allow for more effective
heating. To a lesser extent, this effect is also present in the
effective potential run, but here the faster advection (i.e. shorter
exposure time) of the accreted material through the smaller gain
region around the compact neutron star compensates for the increase in
the local heating rate. On the other hand, the enhancement of the
neutrino heating in the GR model is strong enough to overcome this
competing effect and to shift the balance between neutrino heating and
advection in the gain region far enough to achieve favourable
conditions for the development of an explosion. We emphasize, however,
that despite such differences, the effective potential approximation
provides a remarkable improvement over the purely Newtonian treatment
also in multi-dimensional supernova models.

Incidentally, we also find that the heating conditions depend quite
sensitively on the neutrino microphysics (as already noted by
\citealt{rampp_proc_02} and \citealt{bruenn_10}) as the GR run with
simplified interactions rates fails to develop an explosion. The more
optimistic evolution of the model with the improved rates stems
primarily from the reduction of the $\bar{\nu}_e$ scattering
cross-section on nucleons due to weak magnetism and nucleon
correlations \citep{horowitz_02}, which helps to enhance $\bar{\nu}_e$
luminosities and mean energies (right panel of
Fig.~\ref{fig:dynamics}).

\begin{figure}
\begin{center}
\includegraphics[width=0.45 \linewidth]{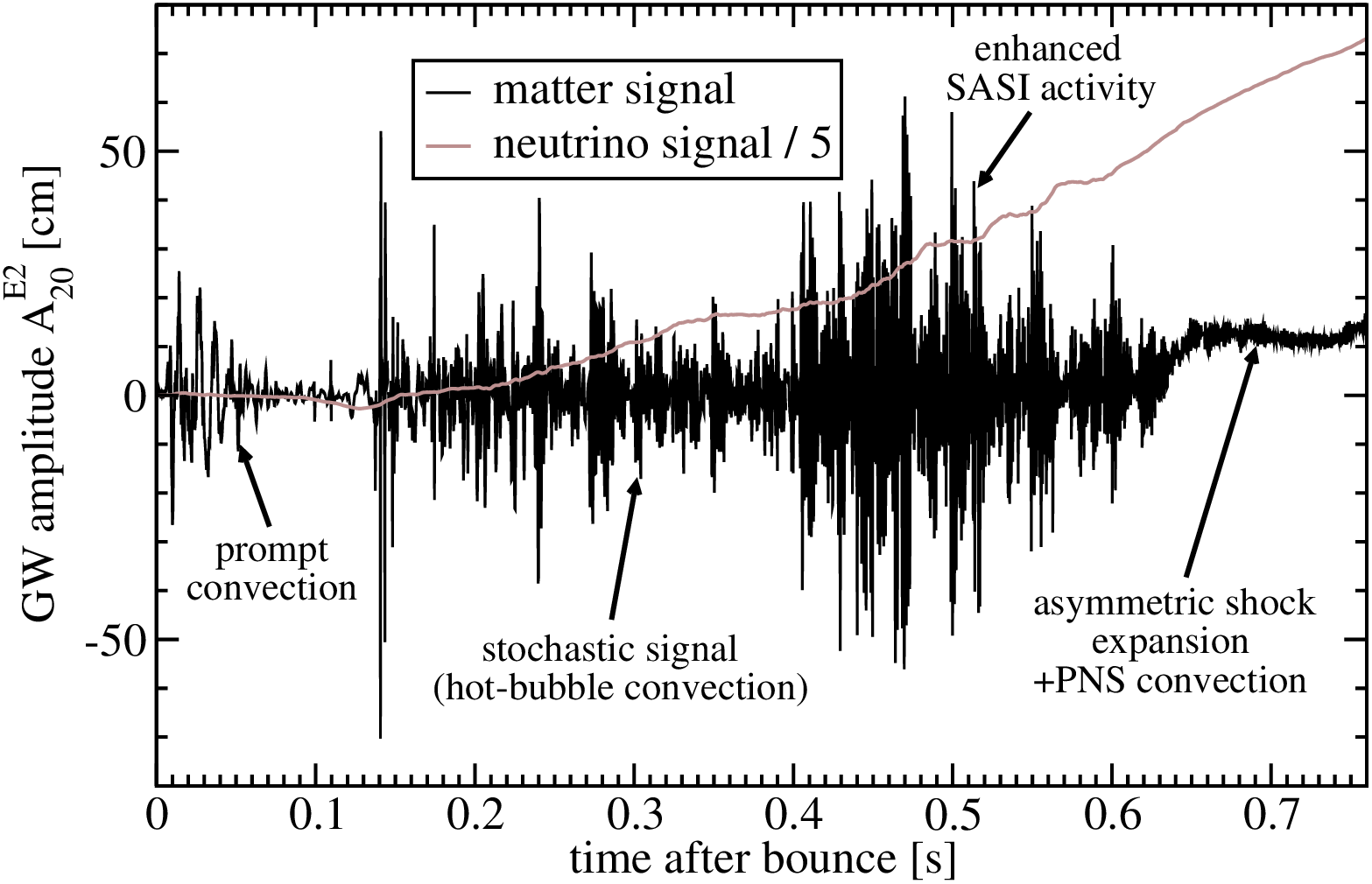}
\includegraphics[width=0.45 \linewidth]{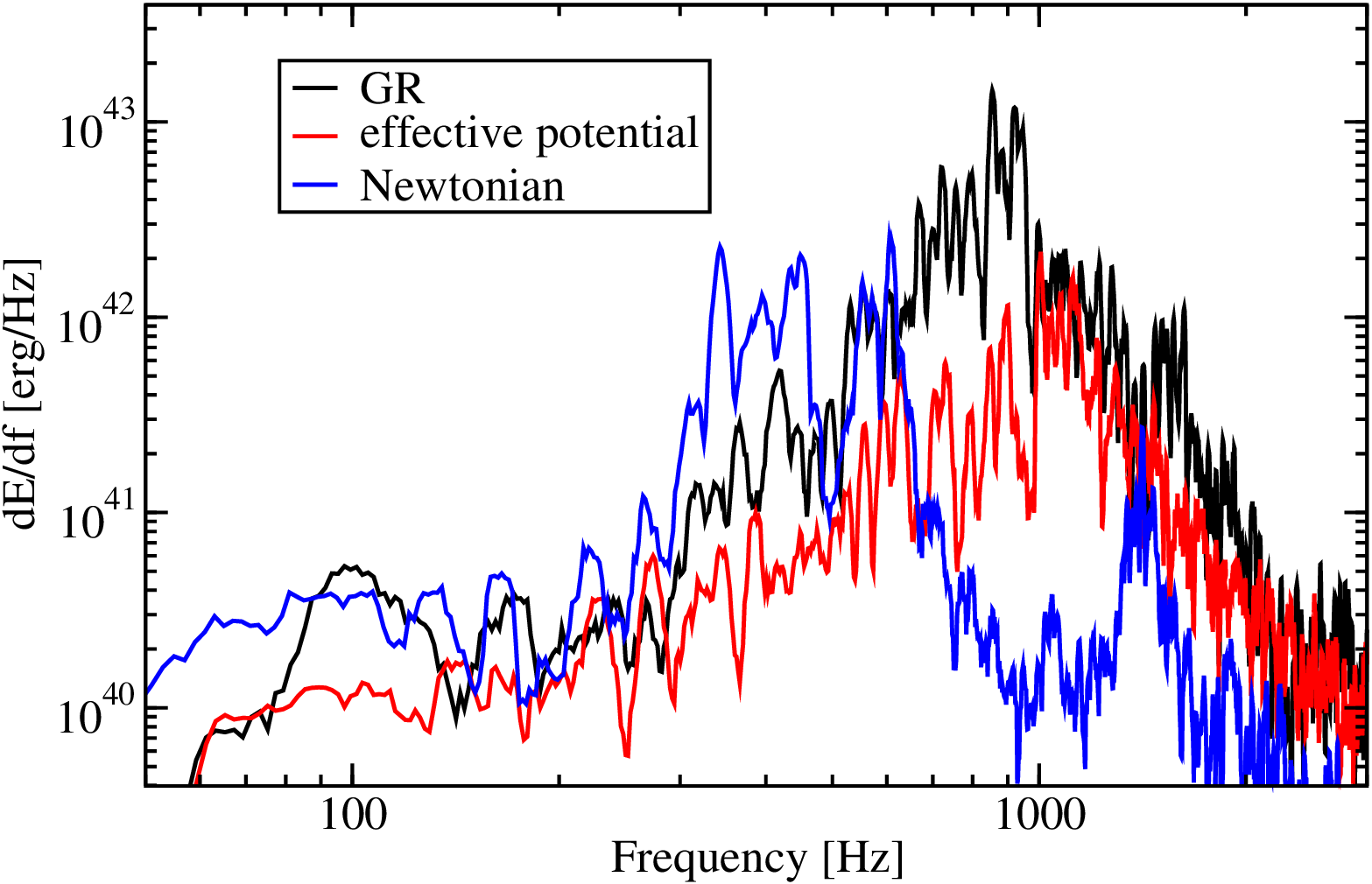}
\caption{Left: Matter (black) and neutrino (light brown) gravitational
  wave signals for the general relativistic $15 M_\odot$ explosion
  model. Right: Gravitational wave spectrum for the first $0.5
  \ \mathrm{s}$ of the post-bounce evolution of a $15 M_\odot$
  progenitor for simulations depending on the treatment of gravity
  (black: GR hydro, red: Newtonian hydro + effective potential, blue:
  purely Newtonian).
\label{fig:gw}}
\end{center}
\end{figure}

The treatment of GR also turns out to be a crucial factor for the
prediction of the gravitational wave signal. While the wave signal for
our relativistic explosion models qualitatively shows the typical
features known from (pseudo-)Newtonian studies with sophisticated or
simplified neutrino transport
\citep{kotake_07,marek_08,murphy_09,yakunin_10} with distinct phases
of gravitational wave emission from prompt post-shock convection,
hot-bubble convection, enhanced SASI sloshing motions, and asymmetric
shock expansion (left panel of Fig.~\ref{fig:gw}), the signal spectrum
is rather sensitive to GR effects (right panel of
Fig.~\ref{fig:gw}). In the purely Newtonian case, the integrated
signal (which is dominated by the contribution from hot-bubble
convection) from the $15 M_\odot$ progenitor peaks at distinctly lower
frequencies around $\sim 500 \ \mathrm{Hz}$ than in the GR case ($\sim
900 \ \mathrm{Hz}$), which is a consequence of a lower
Brunt-{V\"ais\"al\"a} frequency and hence a less rapid deceleration of
convective plumes in the stably stratified cooling region above the
surface of a more extended proto-neutron star surface with lower
surface gravity (cp. the analysis of \citealt{murphy_09}). On the
other hand, the median frequency is somewhat overestimated by the
effective potential approach ($\sim 1100 \ \mathrm{Hz}$) because the
lower neutrino luminosities and mean energies result in a steeper
density stratification in the cooling region, which in turn leads to a
more abrupt braking of convective bubbles. The gravitational wave
signal is thus at least as sensitive to GR effects as to other
physical key parameters such as the equation of state
\citep{marek_08}.

\section{Conclusions and Outlook}
Multi-dimensional general relativistic simulations of core-collapse
with a sophisticated treatment of the microphysics and the neutrino
transport on par with the best currently available Newtonian models
have only recently become possible, but the first available results
presented here in this paper already serve to underscore the
importance of general relativity in the supernova problem. For a $15
M_\odot$ progenitor, we found that GR somewhat improves the heating
conditions compared to models computed in the Newtonian and the
effective potential approximations, which, unlike the GR model, fail to
explode. The gravitational wave spectra are also considerably changed
by GR effects, which shift the typical frequency of the
time-integrated signal upward by $\sim 80\%$ compared to the purely
Newtonian case. We therefore conclude that an accurate treatment of GR
effects may be no less relevant for a better understanding of the
neutrino-driven explosion mechanism and quantitative predictions of the
signals from core-collapse supernovae than other (undoubtedly
important) key factors that have recently been discussed such as
dimensionality issues \citep{nordhaus_10,hanke_11,takiwaki_11} and the
nuclear equation of state.

\acknowledgements This work has been supported by the Deutsche
Forschungsgemeinschaft through the Transregional Collaborative
Research Centres SFB/TR27 ``Neutrinos and Beyond'' and SFB/TR7
``Gravitational Wave Astronomy'' and the Cluster of Excellence EXC-153
``Origin and Structure of the Universe''
(http://www.universe-cluster.de). The simulations discussed here have
been performed on the IBM Power6 p575 system at the Rechenzentrum
Garching (RZG) and the NEC SX-8 at the HLRS Stuttgart.

\bibliography{paper}

\end{document}